\documentclass[epj]{webofc}
\usepackage[varg]{txfonts}   
\usepackage{hyperref}

\renewcommand*{\eqref}[1]{Eq.~(\ref{eq:#1})}

\newcommand*{\seclab}[1]{\label{sec:#1}}

\usepackage{subfigure}

\wocname{ARENA-2016}

\woctitle{ARENA-2016}

\begin{document}

\title{Cosmogenic photons strongly constrain UHECR source models}

\author{\firstname{Arjen} \lastname{van Vliet}\inst{1}\fnsep\thanks{\email{a.vanvliet@astro.ru.nl}}}

\institute{Department of Astrophysics/IMAPP, Radboud University, P.O. Box 9010,
6500 GL Nijmegen, The Netherlands.
          }

\abstract{
With the newest version of our Monte Carlo code for ultra-high-energy cosmic ray (UHECR) propagation, CRPropa 3, the flux of neutrinos and photons due to interactions of UHECRs with extragalactic background light can be predicted. Together with the recently updated data for the isotropic diffuse gamma-ray background (IGRB) by Fermi LAT, it is now possible to severely constrain UHECR source models. The evolution of the UHECR sources especially plays an important role in the determination of the expected secondary photon spectrum. Pure proton UHECR models are already strongly constrained, primarily by the highest energy bins of Fermi LAT's IGRB, as long as their number density is not strongly peaked at recent times. 
}

\maketitle

\section{Introduction}
\label{intro}

Recently the Fermi-LAT collaboration updated their measurements on the isotropic diffuse gamma-ray background (IGRB) and extended it up to 820 GeV~\cite{Ackermann:2014usa}. A possible source for part of the IGRB is secondary electromagnetic cascades initiated by interactions of ultra-high-energy cosmic rays (UHECRs) with the cosmic microwave background (CMB) or the extragalactic background light (EBL). In these same interactions secondary neutrinos can be produced. These neutrinos could possibly contribute to the astrophysical neutrino flux as measured by IceCube~\cite{Aartsen:2015zva}.

The UHECR energy spectrum has been measured with unprecedented statistics by the Pierre Auger~\cite{Aab:2015bza,ThePierreAuger:2015rha} (Auger) and Telescope Array~\cite{Jui:2015tac} (TA) collaborations. The UHECR mass measured by these two collaborations can, however, still be interpreted in different ways.  While the measurements of Auger show a depth of the shower maximum, $X_{\mathrm{max}}$, indicating an increasingly heavier mass composition for $E \gtrsim10^{18.3}$~eV~\cite{Porcelli:2015pac}, TA results in the same energy range are consistent with a pure proton composition~\cite{Fujii:2015tac}. Despite these differences the $X_{\mathrm{max}}$ measurements are in good agreement with each other~\cite{Unger:2015ptc}. The predictions of different air shower simulation models, however, leave room for varying interpretations of the data. Therefore many UHECR composition models are still viable.

However, as shown e.g. in Refs.~\cite{Gavish:2016tfl,Berezinsky:2016jys,Supanitsky:2016gke}, the parameter range of possible pure proton models can be constrained when taking into account the secondary gamma-ray and neutrino production during the propagation of UHECRs from their sources to Earth. Ref.~\cite{Heinze:2015hhp} even claims that the proton dip model is challenged at more than $95\%$ C.L. by the cosmogenic neutrino flux alone. 

To obtain the predicted cosmogenic gamma-ray and neutrino flux for a certain UHECR model the propagation of UHECRs through the universe, including all relevant interactions with the CMB and EBL, has to be simulated. Here the newest version of our UHECR propagation code, CRPropa version 3~\cite{Batista:2016yrx}, is used to simulate the cosmic ray, electromagnetic cascade and neutrino propagation and obtain predictions for the cosmogenic gamma-ray and neutrino fluxes. CRPropa is a full simulation framework for Monte Carlo UHECR propagation including all relevant interactions for protons as well as for heavier nuclei (photo-meson production, pair production, photodisintegration, nuclear decay and energy reduction due to the adiabatic expansion of the universe). For the electromagnetic cascade propagation the specialized code DINT~\cite{Lee:1996fp}, interfaced and shipped with CRPropa 3, is used. DINT solves the one-dimensional transport equations for electromagnetic cascades initiated by electrons, positrons or photons and includes single, double and triplet pair production, inverse-Compton scattering and synchrotron radiation. 

\section{Simulation setup}
\seclab{SimSetup}

All the simulations done here assume a homogeneous distribution of identical sources. The cosmic rays are injected at the sources following a spectrum of
\begin{equation}\label{eq:PowerLawInjection}
	\frac{\text{d}N}{\text{d}E} \propto	(E/E_0)^{-\alpha}\exp(-E/E_{\text{cut}})~,
\end{equation}
with $E$ the energy of the particles, $E_0$ an arbitrary normalization energy, $\alpha$ the spectral index at injection and $E_{\text{cut}}$ the cutoff energy. A minimum energy of cosmic rays at the sources of $E_{\text{min}} = 0.1$~EeV is assumed. The EBL used for the cosmic ray simulations is the Gilmore 2012 model~\cite{Gilmore:2011ks}. In Fig.~\ref{fig:Ev} (solid lines) a reference scenario is given with pure proton injection at the sources for which $\alpha = 2.5$ and $E_{\text{cut}} = 200$~EeV. In this case a co-moving source evolution up to a maximum redshift of $z_{\text{max}}=6$ is implemented. Unless stated otherwise these simulation parameters have been used for the other scenarios as well. 

These parameters are not optimized to fit the cosmic ray spectrum perfectly as the systematic uncertainty of the energy spectrum measurements is much larger than the statistical uncertainty for both Auger and TA. Furthermore the assumption made here, a homogeneous distribution of identical pure proton sources with a power law injection spectrum with exponential cutoff, may not correctly depict the real situation. Additionally it is convenient to change only one parameter at a time in order to see the effect of only that parameter on the predicted spectra. Fitting a pure proton model to the TA spectrum would lead to a strong source evolution and a relatively hard injection spectrum and would be strongly constrained by both the Fermi-LAT IGRB data and the IceCube data~\cite{Heinze:2015hhp,Supanitsky:2016gke}.

\section{Source evolution dependence}
\seclab{EvDep}

\begin{figure}
	\subfigure[Cosmic rays]{
    	\includegraphics[width=.317\textwidth]{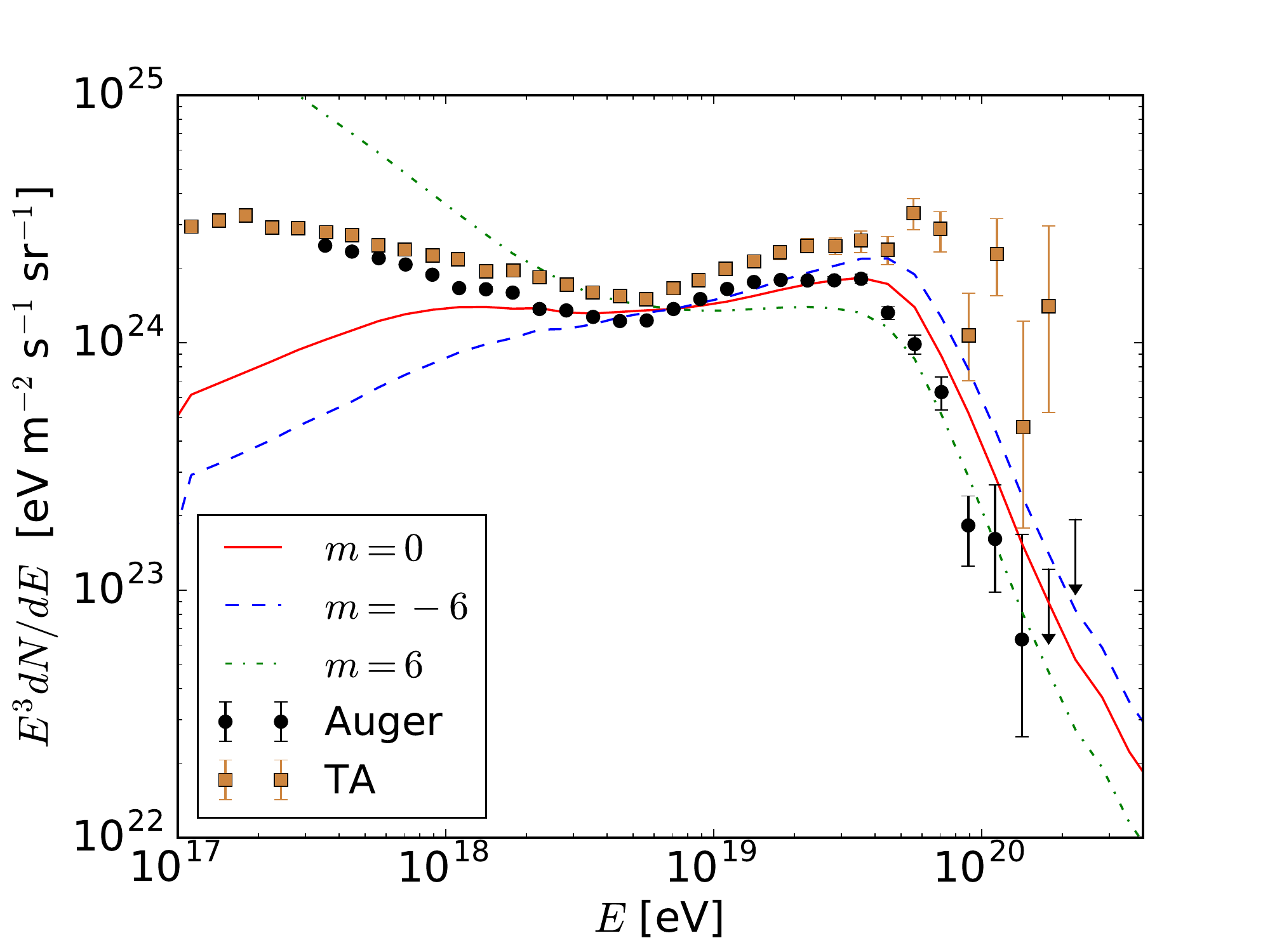}\label{fig:EvCRs}
  	}
  	\subfigure[Neutrinos]{
    	\includegraphics[width=.317\textwidth]{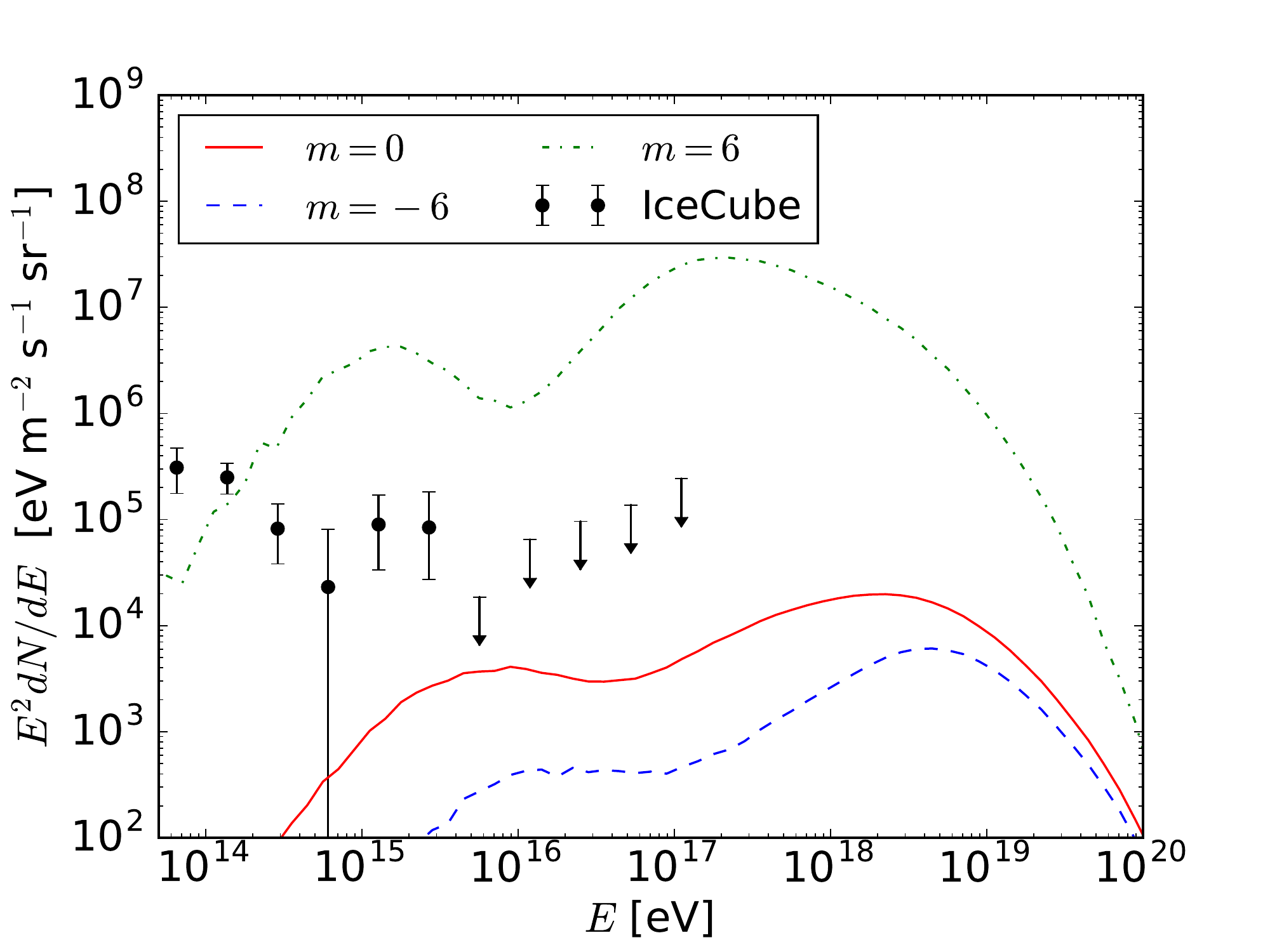}\label{fig:EvNeutrinos}
  	}
  	\subfigure[Photons]{
    	\includegraphics[width=.317\textwidth]{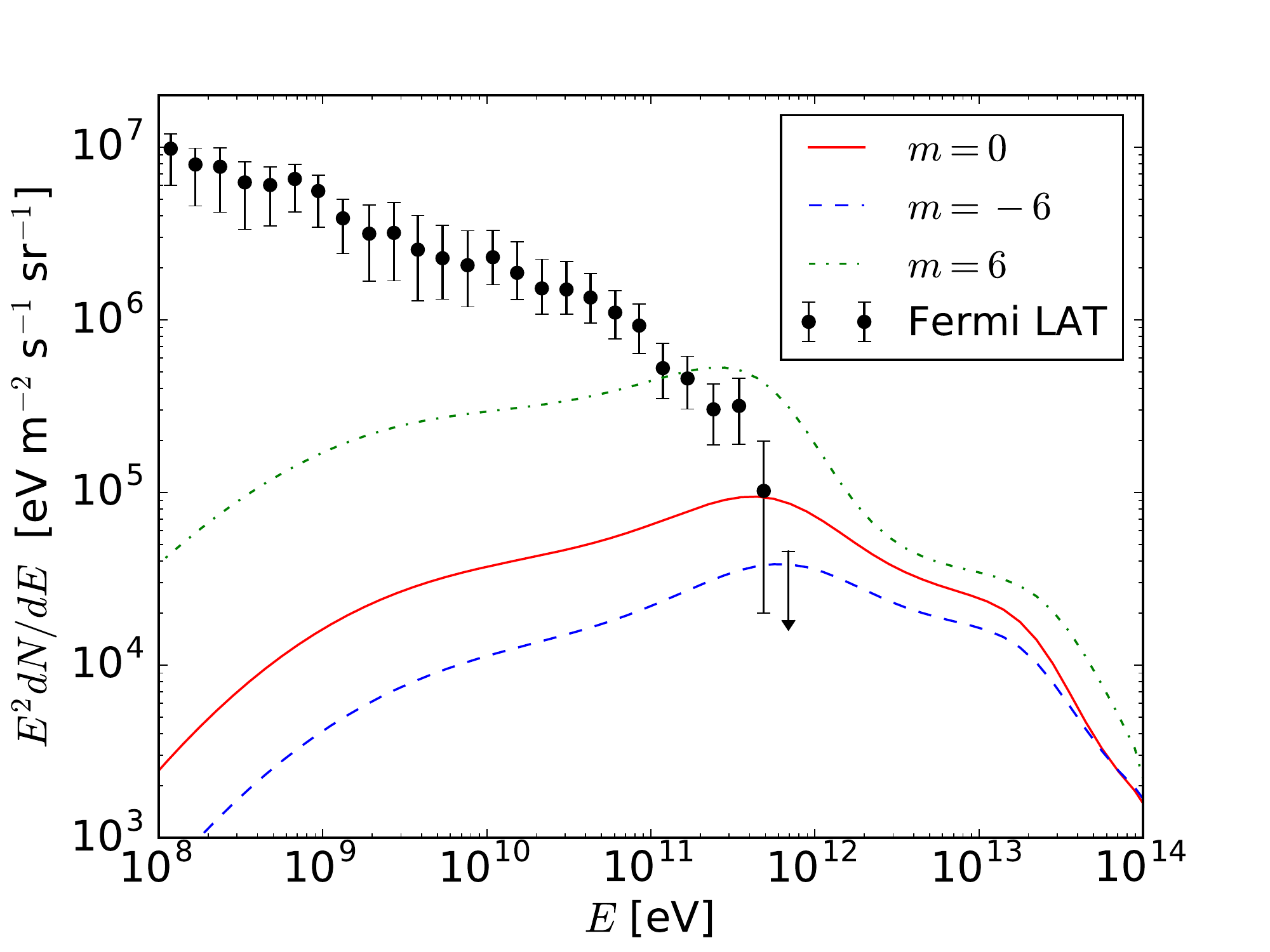}\label{fig:EvPhotons}
  	}
	\caption{ (a) Simulated cosmic ray spectra normalized to the Auger flux~\cite{Aab:2015bza} (circles) at $E = 10^{18.85}$~eV. The spectrum measured by TA~\cite{Jui:2015tac} (squares) is given as well. (b) Corresponding neutrino spectra compared with IceCube data~\cite{Aartsen:2015zva} (circles). (c) Corresponding photon spectra compared with Fermi-LAT IGRB data~\cite{Ackermann:2014usa} (circles) using Galactic foreground model A. The cosmic ray simulations were done starting a pure proton injection at $E_{\text{min}} = 0.1$~EeV with an exponential cutoff at a cutoff energy of $E_{\text{cut}} = 200$~EeV and an injection spectral index of $\alpha = 2.5$ using the Gilmore 2012 EBL model~\cite{Gilmore:2011ks}. The simulations were done up to a maximum redshift of $z_{\text{max}}=6$ with a comoving source evolution multiplied by $(1+z)^m$ with $m = 0$ (solid lines), $m = -6$ (dashed lines) and $m = 6$ (dashed-dotted lines). See text for further details.}
	\label{fig:Ev}
\end{figure}

The evolution in time (redshift) of the sources of UHECRs influences the expected cosmogenic gamma-ray and neutrino flux significantly. As the sources of UHECRs are unknown, a wide range of possible source evolutions could be applicable. Here the effect of the source evolution on the UHECR spectrum and the expected cosmogenic neutrino and gamma-ray fluxes is investigated by taking the reference scenario and adjusting the source evolution by a factor of $(1+z)^m$ with $z$ the redshift of the source and $-6 \leq m \leq 6$. A value of $m = -6$ (negative source evolution) would, for instance, correspond roughly to the source evolution of High Synchrotron Peaked (HSP) BL Lacs~\cite{Gavish:2016tfl}. All the other parameters of the simulation are kept the same in order to clearly see the effect of the source evolution alone. In Fig.~\ref{fig:Ev} the results are given for $m = 0$, $m = -6$ and $m = 6$. The simulated cosmic ray spectrum is normalized to the flux measured by Auger~\cite{Aab:2015bza} at $E = 10^{18.85}$~eV and the simulated cosmogenic neutrino and gamma-ray spectra are normalized accordingly. For comparison the spectrum measured by TA~\cite{Jui:2015tac} is given as well. The simulated neutrino flux is compared with IceCube data~\cite{Aartsen:2015zva} while the simulated gamma-ray flux is compared with Fermi-LAT IGRB data~\cite{Ackermann:2014usa}, using galactic foreground model A with foreground model uncertainties and IGRB intensity uncertainties added in quadrature.

From Fig.~\ref{fig:EvCRs} can be seen that the cosmic ray spectrum is affected by the source evolution most strongly at the lower energy range ($E \lesssim 5\times10^{18}$~eV). This is expected as in this energy range the particles can reach us from very large distances while at higher energy, due to the different energy-loss processes, the cosmic rays can only come from relatively nearby sources. Note here that this lower energy range might also be the regime where a Galactic contribution to the cosmic ray spectrum starts playing a role, which could possibly compensate for an underprediction of the cosmic ray spectrum. Fig.~\ref{fig:EvNeutrinos} shows that a strong source evolution is clearly constrained by the astrophysical neutrino flux as measured by IceCube, but a co-moving or negative source evolution are still allowed when only the neutrino flux is considered. However,  Fig.~\ref{fig:EvPhotons} indicates that especially the non-detection at the highest energy bin (580-820 GeV) of the IGRB measured by Fermi LAT is very constraining. Only sources with a number density strongly peaked at recent times are still allowed by this limit. Note furthermore that there are even many other sources expected to contribute to the IGRB (see e.g. Ref.~\cite{TheFermi-LAT:2015ykq}), which are not taken into account here.

\begin{figure}
	\subfigure[Cosmic rays]{
    	\includegraphics[width=.317\textwidth]{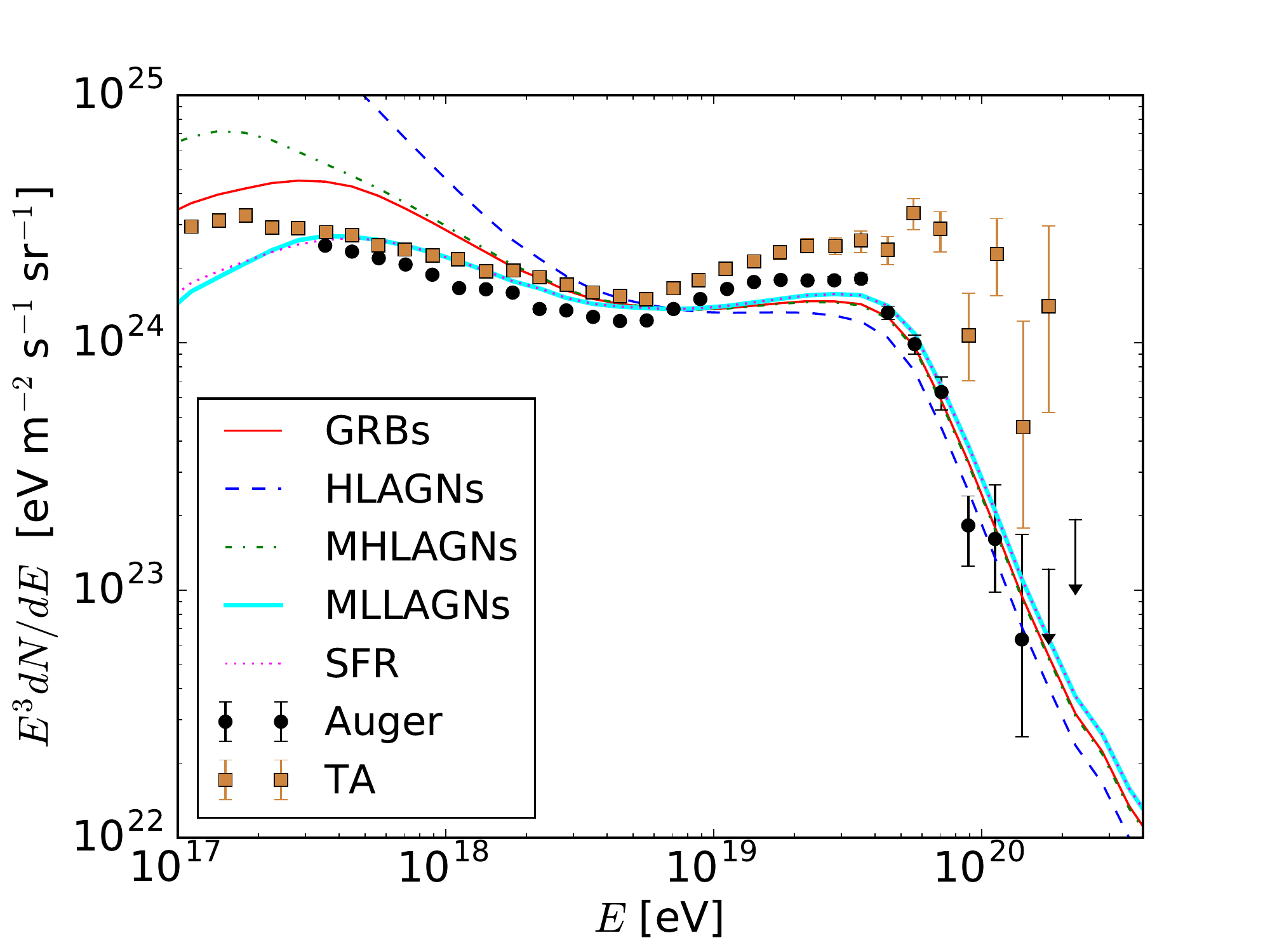}\label{fig:EvSpecCRs}
  	}
  	\subfigure[Neutrinos]{
    	\includegraphics[width=.317\textwidth]{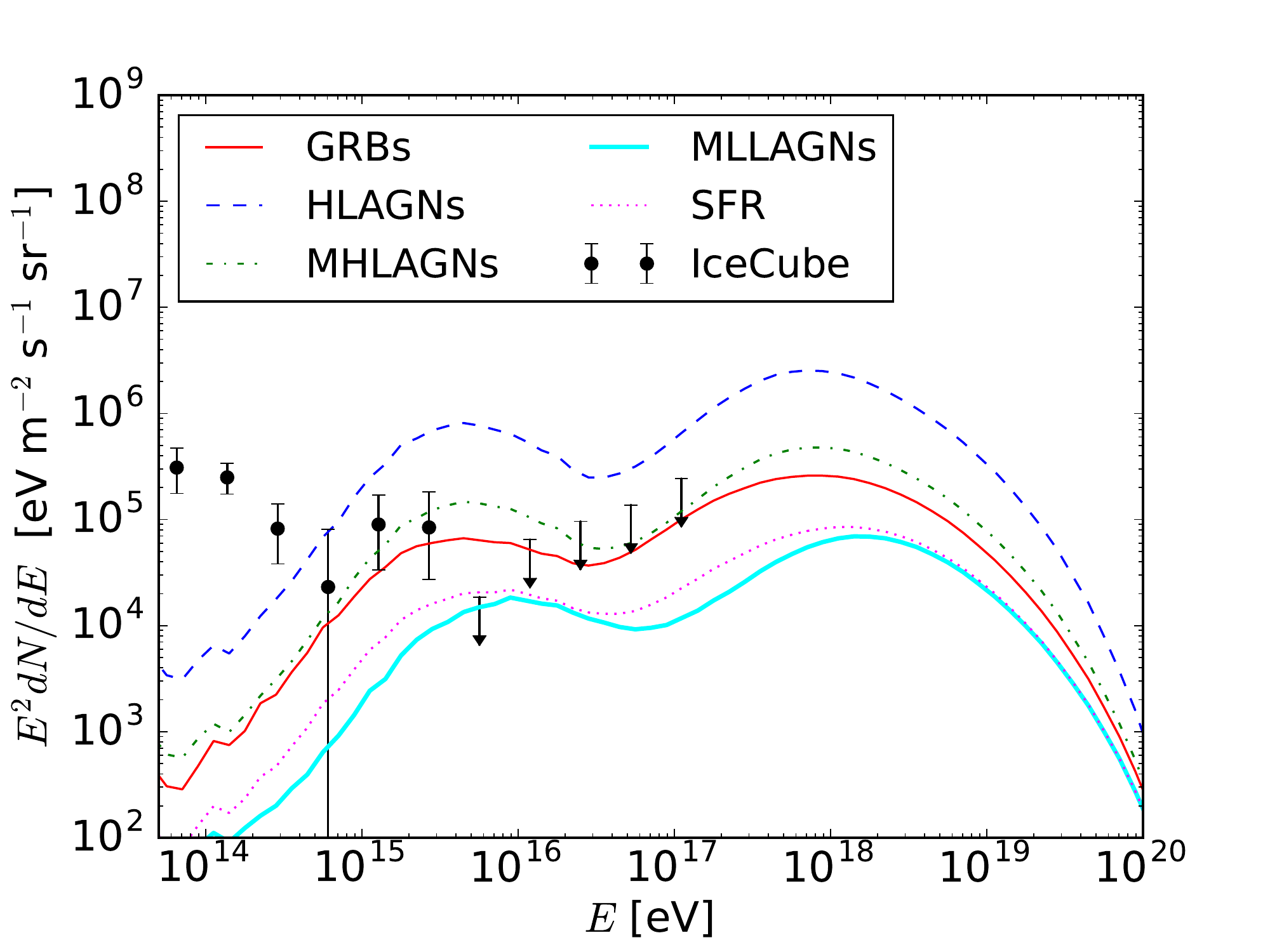}\label{fig:EvSpecNeutrinos}
  	}
  	\subfigure[Photons]{
    	\includegraphics[width=.317\textwidth]{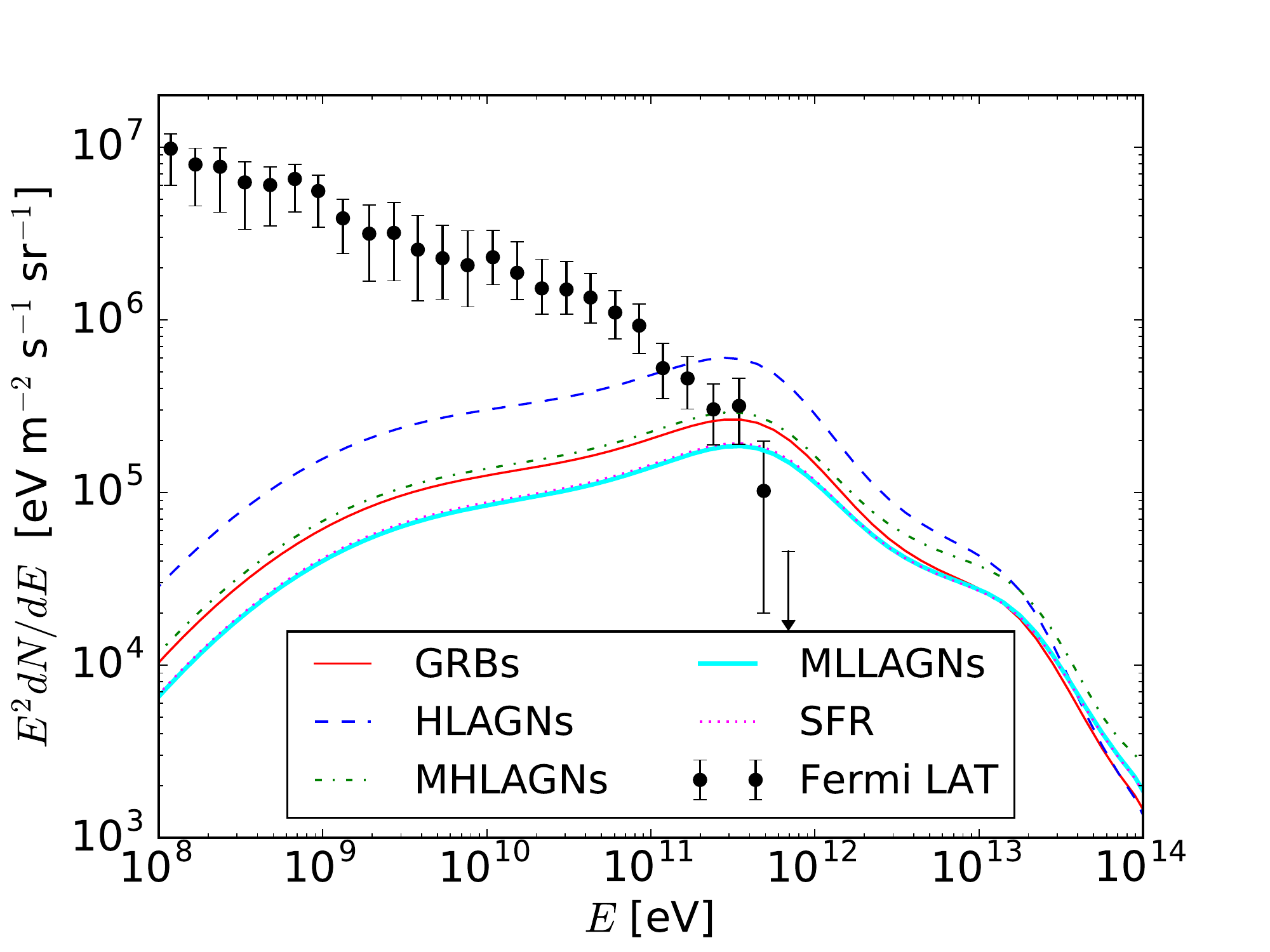}\label{fig:EvSpecPhotons}
  	}
	\caption{ Simulated cosmic ray (a), neutrino (b) and photon (c) spectra compared with the same data as in Fig.~\ref{fig:Ev}. The evolution of the UHECR sources are in this case for GRBs (solid lines), HLAGNs (dashed lines), MHLAGNs (dashed-dotted lines), MLLAGNs (thick solid lines) and sources following the SFR (dotted lines).  See text for further details.}
	\label{fig:EvSpec}
\end{figure}

Instead of a co-moving source evolution multiplied by $(1+z)^m$, specific functions for certain UHECR source candidates can be tested in the same way. For this procedure the source evolution parametrizations listed in Ref.~\cite{Gavish:2016tfl} for gamma-ray bursts (GRBs), high luminosity active galactic nuclei (HLAGNs), medium high luminosity AGNS (MHLAGNs), medium low luminosity AGNs (MLLAGNs) and sources following the star formation rate (SFR) are implemented. The results (with all other parameters of the simulations kept the same) are shown in Fig.~\ref{fig:EvSpec}. From Fig.~\ref{fig:EvSpecPhotons} can be seen that all these source classes exceed the IGRB data measured by Fermi LAT.

\section{Conclusions}
\seclab{Conclusions}

Figs.~\ref{fig:Ev} and~\ref{fig:EvSpec} show that the IGRB measured by Fermi LAT is more constraining for UHECR models than the IceCube neutrino measurements.  The bin with the highest energy (580-820 GeV) of the IGRB especially provides a strong constraint for pure proton UHECR models. For the scenarios investigated here, only in the case of source densities that are strongly decreasing with redshift (for instance in the case of HSP BL Lacs) is it possible to get a gamma-ray spectrum in agreement with that highest energy bin. With a few more years of data and, perhaps, an extension to even higher energies, Fermi LAT might be able to rule out all realistic UHECR pure proton models.

\begin{acknowledgement}

I want to thank Rafael Alves Batista and J\"org H\"orandel for helpful discussions and the Auger PC for their suggested corrections.
I acknowledge financial support from the NWO Astroparticle Physics grant WARP.

\end{acknowledgement}

\end{document}